\documentclass[pra,reprint,amsfonts,amsmath,amssymb,showpacs,twocolumn]{revtex4}

\usepackage[dvipdfmx]{graphicx}
\usepackage{dcolumn}
\usepackage{bm}
\usepackage{amsthm}
\usepackage{mathrsfs}
\usepackage{comment}
\usepackage{color}

\DeclareMathOperator{\diag}{diag}

\DeclareMathOperator{\aut}{Aut}

\begin{document}
\title{Topological influence and back-action between topological excitations}
\author{Shingo Kobayashi$^1$}
\altaffiliation{Present Address: Department of Applied Physics, Nagoya University, Nagoya 464-8603, Japan}
\author{Nicolas Tarantino$^2$}
\author{Masahito Ueda$^1$}
\affiliation{$^1$Department of Physics, University of Tokyo, 7-3-1, Hongo, Bunkyo-ku, Tokyo 113-0033, Japan \\ $^2$Physics and Astronomy Department, Stony Brook University, Stony Brook, NY 11794-3800 USA}

\date{\today}

\begin{abstract}
Topological objects can influence each other if the underlying homotopy groups are non-Abelian. Under such circumstances, the topological charge of each individual object is no longer a conserved quantity and can be transformed to each other. Yet, we can identify the conservation law by considering the back-action of topological influence. We develop a general theory of topological influence and back-action based on the commutators of the underlying homotopy groups. In the case of the topological influence of a half-quantum vortex on a point defect, we point out that the topological back-action from the point defect is a twisting of the vortex. The total twist of the vortex line compensates for the change in the point-defect charge to conserve the total charge. We use this theory to classify charge transfers in condensed matter systems and show that a non-Abelian charge transfer can be realized in a spin-2 Bose-Einstein condensate.    
\end{abstract}
\maketitle
\section{Introduction}
\label{sec:intro}
Topological excitations are the hallmark of symmetry broken systems, such as liquid crystals~\cite{Gennes:1993,Kleman:1983}, superfluid helium systems~\cite{Donnelly,Vollhardt:1990,Volovik:2003}, and ultracold atomic gases~\cite{Stamper-Kurn:2012,Kawaguchi:2012}. They appear as spatial variations of order parameters with a topological charge. Examples include a vortex (line defect), a point defect, and a skyrmion. The topological excitations arise as a result of the spontaneous symmetry breaking of the system and their existence is determined by the broken symmetry. To characterize the broken symmetry of a given order parameter $\psi$, we introduce a symmetry group $G$ whose action $\psi \to g \psi$ $(^{\forall}g \in G)$ does not change the energy of system. Also, its subgroup $H \subset G$ keeps the order parameter invariant, where $H$ is called an isotropy group $H=\{g \in G | g \psi = \psi \}$. In terms of $G$ and $H$, the broken symmetry is described by a quotient space $G/H$, which represents the degeneracy of the order parameter and defines an order parameter manifold (OPM).      

Topological excitations are classified by the homotopy group~\cite{Kleman:1983,Volovik:2003,Mermin:1979,Michel:1980,Trebin:1982,Mineev:1998}. The invariance of a topological charge implies that a topological excitation cannot continuously transform to an other topological excitation if their topological charges are different. For the case of a vortex, the topological charge is characterized by the fundamental group $\pi_1 (G/H)$.  For general cases, in $d$-dimensional ($d$D) space, a $\nu$-dimensional topological excitation is classified by $\pi_{d-\nu-1} (G/H)$. In particular, for $d=3$, topological excitations with the dimension $\nu=0,1,2$ correspond to point defects, line defects, and domain walls, respectively. For example, in the case of a scalar Bose-Einstein condensate (BEC) or a superfluid $^4$He, the OPM is $U(1)$, which represents the global gauge degrees of freedom. The fundamental group $\pi_1 (U(1)) \cong \mathbb{Z}$ gives a topological charge of a point defect in 2D space ($d=2,\nu=0$) or that of a line defect in 3D space ($d=3,\nu=1$), where $\mathbb{Z}$ is an additive group of integers.  

Topological excitations are usually classified by the homotopy group $\pi_{n} (G/H)$, but we cannot, in general, directly apply this scheme if more than one type of topological excitations coexist, for there may be a topological influence between them. The topological influence was first pointed out by Mineev and Volovik \cite{Volovik:1977inf}, and Mermin \cite{Mermin:1978inf} in condensed matter physics, and Schwarz \cite{Schwarz:1982} in cosmology. For example, a point-defect charge changes its sign as it makes a complete circuit of a half-quantum vortex. The topological influence depends only on the topology of the order parameter manifold and is mathematically described by the $\pi_{1} (G/H)$ action on $\pi_{n} (G/H)$. That is, the homotopy groups no longer uniquely determine the charge of the topological excitation under the topological influence because the charge is not invariant under the relative motion of one topological object against another. In a previous work~\cite{Kobayashi:2012}, we have demonstrated that the Abe homotopy group $\kappa_{n} (G/H)$ can be used to deal with the topological influence. The Abe homotopy group classifies maps from a pinched torus (see Fig. 6 in Ref.~\cite{Kobayashi:2012}) to the order parameter manifold, and this map enables us to classify a combined object of line defects and point defects. In this framework, the $\pi_{1} (G/H)$ action on $\pi_n (G/H)$ can naturally be described by the group action of $\kappa_{n} (G/H)$.

So far, these researches focus on the topological influence in multiple topological excitation systems, but do not address the question of the charge conservation. In the present paper, we address the consistency between the topological influence and the charge conservation. The charge conservation is important for an isolated system like ultracold atomic gases because a topological charge cannot escape from the system. The consistency between topological influence and the charge conservation was first discussed by Bucher {\it et al}.~\cite{Bucher:1992} in connection with Alice cosmic rings. They discussed a topological charge transfer from the magnetic monopole to the Alice cosmic ring. The purpose of this paper is to revisit the relation between the charge conservation and the charge transfer in isolated systems and to show that the back-action from a point defect can be interpreted as a ``twist of a vortex". We formulate the charge transfer using the commutator of the underlying homotopy groups $[\pi_1, \pi_n]$, which gives a general classification of the charge transfer. Here, an element of $[\pi_1, \pi_n]$ lies within $\pi_n (G/H)$ even for $n \ge 2$ because the $\pi_1 (G/H)$ action on $\pi_n (G/H)$ is thought of as a homomorphism from $\pi_1(G/H)$ to $\aut(\pi_n)$ which is the automorphism map on $\pi_n(G/H)$~\cite{Switzer:1975}. We also make concrete classifications of the charge transfer for liquid crystals, ultracold atomic gases, and superfluid $^3$He and the non-Abelian charge transfer in the spin-2 BEC.          

This paper is organized as follows. In Sec.~\ref{sec:influence}, we review a topological influence using liquid crystals, which exhibit the topological influence on non-Abelian vortices for a biaxial nematic phase and on a point defect in a uniaxial nematic phase. In general, the topological influence is described by the automorphism map on $\pi_n$ induced by $\pi_1$. In Sec.~\ref{sec:twist}, we describe the connection between the back-action and the twist of vortex. In Secs.~\ref{sec:transfer-vortex}-\ref{sec:transfer-monopole}, we discuss a charge transfer between non-Abelian vortices and that between a vortex and a point defect. In Sec.~\ref{sec:examples}, we gives examples in liquid crystals, spinor BECs, and superfluid $^3$He. In Sec.~\ref{sec:testing}, we suggest the experimental detection of the total charge and each individual topological charge under the charge conservation. Finally, in Sec. \ref{sec:summary}, we summarize the main results of this paper. In Appendix, we give the mathematical definitions of the semidirect product and the $h$-product to make this paper self-contained.

\section{Review of The Topological influence between topological charges}
\label{sec:influence}

In this section, we review the topological influence, which is defined as a nontrivial effect on a topological charge of a topological excitation due to the presence of another topological excitation. It is well-known that the topological influence exists between non-Abelian vortices \cite{Kleman:1983,Mermin:1979} and between a point defect and a vortex~\cite{Volovik:1977inf,Mermin:1978inf}. This effect is mathematically described as the $\pi_1 (G/H)$ action on $\pi_n(G/H)$, which makes up an automorphism map on $\pi_n (G/H)$. Especially, for $n=1$, the topological influence is described by a conjugation in group theory. Recently, we have proposed that the Abe homotopy group $\kappa_n (G/H)$~\cite{Kobayashi:2012} provides a mathematical framework for simultaneously classifying a vortex and a topological excitation with dimensionality $n$. This group naturally includes the topological influence in its group structure. In condensed matter systems, the topological influence is predicted for a number of ordered phases and for several topological excitations such as a vortex with a non-Abelian charge \cite{Kleman:1983,Mermin:1979,Makela:2003,Makela:2006,Semenoff:2007}, a point defect~\cite{Volovik:1977inf,Mermin:1978inf,Kobayashi:2012}, and a $\pi_4$-texture~\cite{Kobayashi:2012} in 4D space-time.   

First, we discuss the topological influence on a non-Abelian vortex. As a concrete example, let us consider a biaxial nematic phase in a liquid crystal. An order parameter of the liquid crystal is described by a real traceless symmetric tensor $Q=\sum_{i,j=1,2,3} Q_{ij} \bm{d}_i \otimes \bm{d}_j$. The symmetry of this system is given by $SO(3)$. We define the group action on $Q$ as 
\begin{align}
Q \mapsto RQR^T,
\end{align}
where $R \in SO(3)$ and $T$ means the transpose.

\begin{figure}[tbp]
\centering
 \includegraphics[width=8cm]{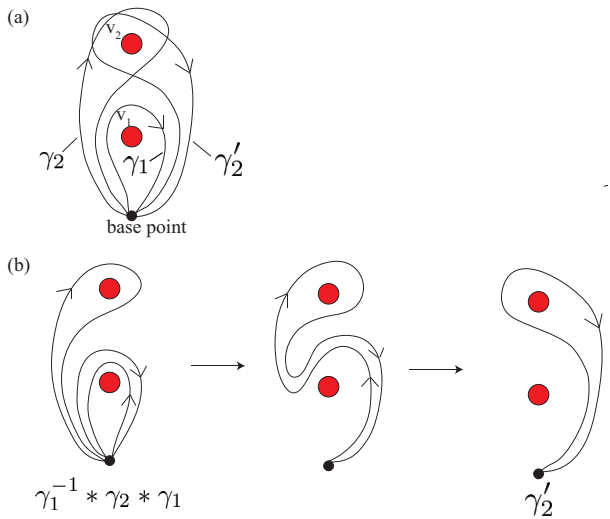}
 \caption{(a) (color online) Schematic illustration of loops enclosing vortices. $\gamma_1$ represents a loop enclosing a vortex v$_1$, and $\gamma_2 (\gamma_2')$ encloses a vortex v$_2$ from the left(right)-hand side of v$_1$, but does not enclose v$_1$. (b) Loop deformation from $\gamma_1^{-1} \ast \gamma_2 \ast \gamma_1$  to $\gamma_2'$, where $\ast$ is a loop product defined by Eq.~(\ref{eq:loop-prod}) and $\gamma_1^{-1}$ represents a loop in the reverse direction. This deformation shows that the loops $\gamma_1^{-1} \ast \gamma_2 \ast \gamma_1$ and $\gamma_2'$ are homotopy equivalent.}\label{fig:loop-deformation}
\end{figure}

In the biaxial nematic phase, the matrix order parameter is described by 
\begin{align}
 Q_{\rm BN} = \diag (A_1,A_2,A_3),
\end{align}
where $A_1,A_2,A_2 \in \mathbb{R}$ and they satisfy $A_1+A_2+A_3=0$. In addition, we assume that $A_1 \neq A_2$, $A_2 \neq A_3$ and $A_3 \neq A_1$. Hence, the isotropy group is the second dihedral group $H_{\rm BN} = D_2 \subset SO(3)$ and the OPM turns out to be \cite{Kleman:1983,Mermin:1979}
\begin{align}
G/H_{\rm BN} = SO(3)/D_2 \cong SU(2)/Q_8,   
\end{align}
where $Q_8$ is the eight-element quaternion group, which is represented in terms of the Pauli matrices $\sigma_i \;(i=x,y,z)$ as $Q_8 := \{ \pm \bm{1}_2 , \pm i \sigma_x , \pm i \sigma_y, \pm i \sigma_z\}$. Here, $\bm{1}_{2}$ is the 2-by-2 identity matrix. Since $SU(2)$ is simply connected, from the exact homotopy sequence, we obtain  
\begin{align}
 \pi_1 (G/H_{\rm BN}) \cong Q_8. \label{eq:vortex_BN}
\end{align}
Therefore, vortices of the biaxial nematic phase are labeled by $Q_8$. Furthermore, $Q_8$ is a non-Abelian group, so that vortices have non-Abelian charges. Let us consider a situation in which two vortices labeled by $i \sigma_y$ and $i \sigma_x$ coexist and a vortex with charge $i \sigma_y$ rotates around a vortex with charge $i \sigma_x$. To see the topological nature of this situation, we carry out this operation adiabatically so as to guarantee that the order parameters remain in the ground-state manifold during the rotation. From Fig.~\ref{fig:loop-deformation}, we find that the situation remains the same if we replace $\gamma_1$ and $\gamma_2$ with $i \sigma_x$ and $i \sigma_y$, respectively, and identify the topological charge of a vortex v$_2$ after rotating around a different vortex v$_1$ with the topological charge of a vortex enclosed by $\gamma_2'$. As a result, the topological charge $i \sigma_y$ changes as    
\begin{align}
 i \sigma_y \mapsto (i \sigma_x)^{-1} (i \sigma_y) (i \sigma_x)= -i \sigma_y. \label{eq:inf-pi1}
\end{align}
 According to the definition of the eight-element quaternion group, $i \sigma_y$ and $-i \sigma_y$ are different topological charges. Thus, a topological charge of a vortex can transform to an other topological charge by making a complete circuit around a different vortex. In other words, the non-Abelian charge is not uniquely determined up to conjugation. This ambiguity is a well-known property in the classification of vortices \cite{Kleman:1983,Mermin:1979}. In fact, it is known that non-Abelian vortices are classified by conjugacy classes of the fundamental group. The conjugacy class of $Q_8$ is given by
 \begin{align}
  [Q_8] =\{ \{ \bm{1}_2\}, \{-\bm{1}_2\}, \{ \pm i \sigma_x\}, \{ \pm i \sigma_y\}, \{ \pm i \sigma_z\} \},
 \end{align}
where $[\cdots]$ means conjugacy classes. Therefore, we have five different types of vortices in the biaxial nematic phase.
However, the classification based on the conjugacy classes presupposes that the topological charge need not be conserved~\cite{Lo:1993}. A classification that respects the topological charge conservation is the primary subject of this paper, which will be discussed in Sec.~\ref{sec:back-action} in detail.  
    
In general, the topological influence on a non-Abelian vortex is described by the inner automorphism map  
\begin{align}
f_{\gamma_1}: \gamma_2 \mapsto \gamma_2' = \gamma_1^{-1} \ast \gamma_2 \ast \gamma_1,
\end{align}
where $\gamma_{1,2}$ and $\gamma_{1,2}' \in \pi_1(G/H)$, and $\ast$ represents the loop product defined for a parameter $t \in [0,1]$ by 
\begin{align}
 (\gamma_1 \ast \gamma_2) (t) = \begin{cases} \gamma_1 (2t) \ \ &\text{if } 0 \le t \le \frac{1}{2}; \\ \gamma_2 (2t-1) \ \ &\text{if } \frac{1}{2} \le t \le 1, \end{cases} \label{eq:loop-prod}
\end{align}
where $\gamma_{1(2)}$ satisfies $\gamma_{1(2)} (0) = \gamma_{1(2)} (1)$.
The loop product satisfies the group structures through the homotopy equivalence~\cite{Nakahara:2003}. Thus, $f_{\gamma_1}$ is the identity map if $\gamma_1$ is an element of the centralizer $Z(\pi_1) \subset \pi_1$.
\begin{figure*}[tbp]
\centering
 \includegraphics[width=11cm]{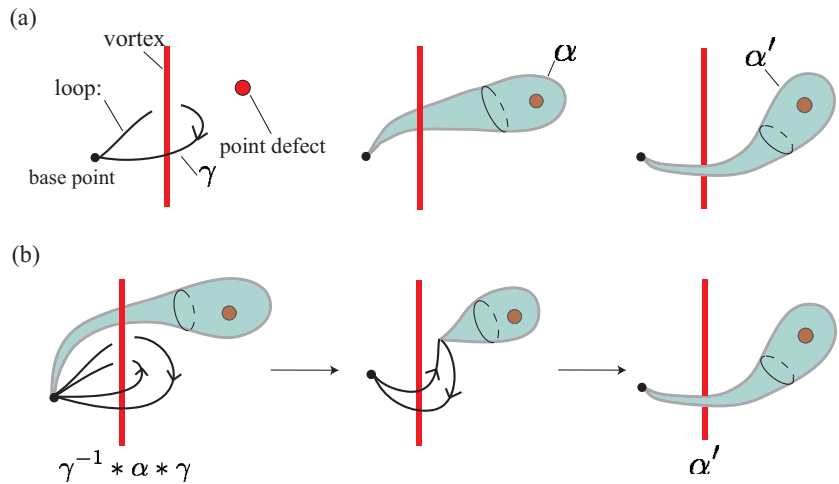}
 \caption{ (a) (color online) Schematic illustration of a loop $\gamma$ enclosing a vortex and a closed surface $\alpha'$ enclosing a point defect in front of the vortex. We define $\alpha$ as a closed surface that encloses the point defect from behind the vortex as shown in the middle figure of (a). Due to the presence of the vortex, $\alpha$ cannot continuously transform to $\alpha'$. In (b), however, by connecting a loop $\gamma$ to $\alpha$ via the loop product (\ref{eq:hiloop-prod}) as $\gamma^{-1} \ast \alpha \ast \gamma$, we can see the homotopy equivalence between $\gamma^{-1} \ast \alpha \ast \gamma$ and $\alpha'$ through continuous deformation.}\label{fig:surface-deformation}
\end{figure*}

Next, let us consider the topological influence on a topological excitation with dimension higher than one ($n \ge 2$). As an example, we treat the topological influence on a point defect in the uniaxial nematic phase in a liquid crystal. An order parameter of the uniaxial nematic phase is given by  
\begin{align}
 Q_{\rm UN} = A \diag (2/3,-1/3,-1/3),
\end{align}
where $A \in \mathbb{R}$. The corresponding isotropy group is $H_{\rm UN} =  SO(2) \rtimes  \mathbb{Z}_2 \cong O(2)$, where ``$\rtimes$" denotes the semidirect product defined in Appendix \ref{app:A}. Thus, the OPM becomes
\begin{align}
G/H_{\rm UN } = SO(3)/(SO(2) \rtimes \mathbb{Z}_2) = S^2 / \mathbb{Z}_2 = \mathbb{R}{\rm P}^2, \label{eq:GH_UN}
\end{align}
where $\mathbb{R}{\rm P}^2$ is a 2D real projective space, whose first and second homotopy groups are given by
\begin{subequations}
\begin{align}
 &\pi_1 (\mathbb{R}{\rm P}^2) \cong \mathbb{Z}_2, \label{eq:un_pi1}\\
 &\pi_2 (\mathbb{R}{\rm P}^2) \cong \mathbb{Z}. \label{eq:un_pi2}
\end{align} 
\end{subequations}
Equations (\ref{eq:un_pi1}) and (\ref{eq:un_pi2}) predict the existence of a vortex and a point defect in the uniaxial nematic phase. Let us denote their elements as $\gamma \in \pi_1(G/H)$ and $\alpha \in \pi_2(G/H)$, and consider a situation in which a vortex with charge $\gamma$ and a point defect with charge $\alpha$ coexist as illustrated in Fig.~\ref{fig:surface-deformation} (a). When the point defect makes a complete circuit around the vortex, the charge of the point defect changes to $\alpha'$. Through continuous transformation as illustrated in Fig.~\ref{fig:surface-deformation} (b), we obtain $\alpha'$ as
\begin{align} 
\alpha' = f_{\gamma} (\alpha) = \gamma^{-1} \ast \alpha \ast \gamma, \label{eq:inf_alpha}
\end{align}
where the loop product $\ast$ is defined as
\begin{align}
 (\gamma \ast \alpha)(t_1,t_2) = \begin{cases} \gamma (2t_1) \ \ &\text{if } 0 \le t_1 \le \frac{1}{2}; \\ \alpha (2t_1-1,t_2) \ \ &\text{if } \frac{1}{2} \le t_1 \le 1. \end{cases} \label{eq:hiloop-prod}
\end{align}
Equation~(\ref{eq:inf_alpha}) represents the topological influence of the vortex on the point defect. By definition, $\alpha' \in \pi_2$ and $f_{\gamma}$ is bijective, so $f_{\gamma}$ acts on the $\pi_2$-element as an automorphism map. Thus, similar to the non-Abelian vortex, the topological influence of the vortex is represented by the automorphism map.  On the other hand, since $\pi_2  \cong \mathbb{Z}$, the set of automorphism maps $\aut (\pi_2)$ is isomorphic to $\mathbb{Z}_2$. As a result, the possible topological influence is given by a $\mathbb{Z}_2$-action on the point-defect charge. 

In the uniaxial nematic phase, a disclination gives the nontrivial $\mathbb{Z}_2$-action. To show this, we stat with the standard order parameter of the uniaxial nematic phase $Q = A \diag (2/3,-1/3,-1/3)$ and describe the order parameters along a loop around the disclination by $R(\pi t,z) Q R (\pi t,z)^T$, where $R(\pi t,z)$ represents a $\pi t$-rotation about the $z$ axis in the Cartesian coordinates and $t \in [0,1]$ is a parameter which specifies the position of the loop. Then, the vector $\bm{d} = (1,0,0)$, which describes the direction of the major axis of the nematic tensor at $t=0$, changes to $-\bm{d}$ by the action of $R(\pi,z)$ at $t=1$. On the other hand, the topological charge of the point-defect is defined by 
\begin{align}
N_2 =  \frac{1}{4 \pi} \iint_{S^2} d \theta d \phi \; \bm{d} \cdot \left( \frac{\partial \bm{d} }{\partial \theta} \times \frac{\partial \bm{d}}{\partial \phi} \right),
\end{align}
where $(\theta, \phi) \in S^2$ are the polar and azimuthal coordinates of a unit sphere enclosing the point defect. The charge $N_2$ transforms into its inverse $-N_2$ if the vector $\bm{d}$ changes to $-\bm{d}$. Thus, when the point defect goes around the disclination, the sign of $N_2$ changes due to the influence of the disclination.

 In general, the topological charge involving the topological influence can be described by the Abe homotopy group $\kappa_n(G/H)$~\cite{Kobayashi:2012}. By analogy with the classification of non-Abelian vortices, the conjugacy class of the Abe homotopy group gives the corresponding topological charge if the charge conservation is not required. The conjugacy class of the second Abe homotopy group in the uniaxial nematic phase is given by~\cite{Kobayashi:2012}
\begin{align}
 [\kappa_2 (G/H)] \cong \mathbb{Z}_2 \times \mathbb{Z}_2.
\end{align}
The first and second $\mathbb{Z}_2$ on the right-hand side represent the charge of a vortex and that of a point defect, respectively. Here the topological influence of the vortex changes the charge of the point defect from $\mathbb{Z}$ to $\mathbb{Z}_2$ in which even and odd integers form two equivalent classes.

\section{Charge conservation and Topological back-action}
\label{sec:back-action}
In this section, we shift focus to the topological phenomena under the charge conservation. Under the topological influence, the total topological charge is {\it prima facie} not conserved. However, in an isolated system such as ultracold atomic gases, the topological charge cannot escape from the system. Thus, the topological influence needs to be revisited so that it is consistent with the charge conservation.  When imposing the charge conservation, we naturally derive the back-action on a vortex as a counteraction of the topological influence. We call it a {\it topological back-action}. The topological back-action is physically defined as a ``twist of the vortex", around which a vortex (a point defect) rotates. In the case of the influence of a vortex, the back-action is equivalent to a change in the topological charge of the vortex at the rotating center. On the other hand, the back-action of a point defect causes two vortex loops on the vortex line at the rotating center. In what follows, we show the relationship between the topological influence and the twist of the vortex, which acquires a nontrivial topological charge given by $\pi_2(G/H)$ when the vortex and the point defect coexist. Furthermore, to characterize the topological influence and the back-action in a unified way, we introduce a {\it topological charge transfer}, which relates the topological influence with the back-action. We formulate the charge transfer based on the scheme of Bucher~{\it at el.}~\cite{Bucher:1992} who discussed the topology of an Alice cosmic ring, which has the same topological nature as that of a half-quantum vortex (HQV) ring in the polar phase of a spin-1 BEC and the superfluid $^3$He-A dipole-free state. Also, we show that the charge transfer can be classified by the commutator $[\pi_1,\pi_n]$ ($n=1,2,\cdots$). 

\begin{figure*}[tbp]
\centering
 \includegraphics[width=12cm]{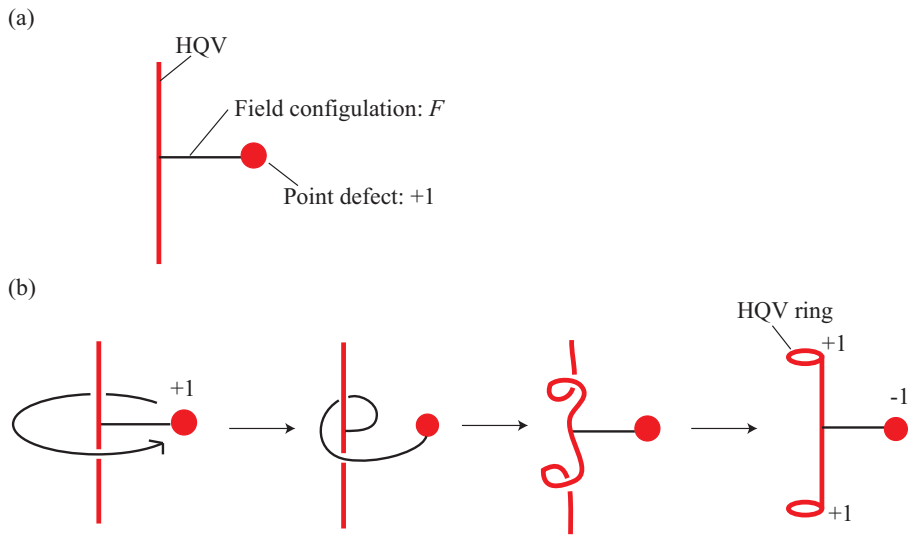}
 \caption{(color online) Schematic illustration showing how the topological back-action leads to the twist of a vortex. (a) A half-quantum vortex (HQV) coexists with a point defect with charge $+1$, where a field configuration line $F$ continuously connects the order parameter around the HQV to that around the point defect. (b) Following the point defect which rotates around the HQV, $F$ winds around the HQV. Then, the HQV twists itself so as to unwind $F$, resulting in a pair of twists along the vortex line, which, in turn, can continuously transform to two HQV rings, each of which carries a $\pi_2$ charge of $+1$. Since the total charge is $2+(-1) = 1$, the charge conservation holds. }\label{fig:back-ation}
\end{figure*}

\subsection{Topological back-action and twist of a vortex}
\label{sec:twist}
First, we show the connection between the topological back-action on a vortex and a twist of a vortex. For the case of non-Abelian vortices, the topological back-action is nothing but a change in the topological charge of the vortex since the topological charges are described by the fundamental group $\pi_1 (G/H)$. On the other hand, for the topological influence on a point defect, the topological back-action on a vortex is no longer described by the fundamental group because these charges belong to the higher homotopy groups. Thus, we call for a new understanding in higher-dimensional ($n \ge 2$) cases. To resolve this problem, we show that a twist of a vortex gives the consistent charge.  

To understand the relation between the topological influence and the twist of a vortex intuitively, let us consider a HQV and specify a field configuration $F$ along a line which connects an order parameter of a point defect with that of the HQV as illustrated in Fig.~\ref{fig:back-ation} (a). As shown in Fig.~\ref{fig:back-ation} (b), if the point defect makes a complete circuit around the HQV, $F$ also winds itself around the HQV once. After that, we twist the HQV so as to unwind $F$. As a result, the winding of $F$ gives rise to the twist of the vortex, which is a direct manifestation of topological back-action. Furthermore, this twist can continuously transform to a pair of HQV rings, each of which carries not only the charge of a vortex, but also that of a point defect~\cite{Mori:1988,Nakanishi:1988,Bais:1995}. Each HQV ring thus carries the topological $\pi_2$ charge of +1 due to the nontrivial twist, so that the total topological charge is invariant under the topological influence such that $0+1 \to 2 + (-1)$. 

In what follows, we prove that the twist of a vortex is caused by the topological back-action, which is associated with the topological charge given by $\pi_2(G/H)$. To characterize a topological charge, we first define a loop space as
\begin{align}
 \Omega(G/H) := \text{Maps}[S^1 \to G/H],
\end{align}   
where Maps[$S^1 \to G/H$] is a set of maps from $S^1$ to $G/H$, which is called a loop space. Each element of $\Omega (G/H)$ corresponds to a loop $l (s_1)$, where $ s_1 \in [0,1]$.  When a loop encloses a vortex core, its homotopy equivalence classes define the topological $\pi_1$ charge of the vortex. When we consider a vortex line, we have a set of loops along the vortex line. Without loss of generality, we can assume that a loop enclosing the top edge of the vortex line $l_{\rm i}(s_1)$ has the same configuration as that enclosing the bottom edge of the vortex line $l_{\rm f}(s_1)$ as long as there exist a continuous transformation between them. To characterize a vortex line, we redefine each loop as $L (s_1,s_2)$, where $s_1 \in [0,1]$ parametrizes each loop and $s_2 \in [0,1]$ parametrizes the location of the loop along a vortex line. For a given $s_2 \in [0,1]$, $L(s_1,s_2)$ corresponds to a loop $l(s_1)$ enclosing a vortex core at $s_2$. By assumption, $L(s_1,s_2)$ satisfies the boundary condition as $L(s_1,0) = l_{\rm i} (s_1) = l_{\rm f} (s_1) = L(s_1,1)$. Thus, a whole set of loops described by a configuration of a vortex line is characterized by $L(s_1,s_2)$. This vortex line can transform to another vortex line characterized by $L'(s_1,s_2)$ if $L(s_1,s_2)$ is homotopic to $L'(s_1,s_2)$ within $G/H$. This statement is equivalent to the classification of loops in the loop space $\Omega (G/H)$ so that $L(s_1,s_2)$ represents a loop in $\Omega (G/H)$. Thus, a topological charge of the vortex line is given by the fundamental group of the loop space: $\pi_1 (\Omega (G/H))$. For a loop space, the following isomorphism holds (e.g. see Ref.~\cite{Switzer:1975} chapter 1):
\begin{align}
 \pi_1 (\Omega (G/H)) \cong \pi_2 (G/H).
\end{align}   
Thus, a nontrivial twist of a vortex line is equivalent to a topological charge classified by $\pi_2 (G/H)$. 

\subsection{Charge transfer between two non-Abelian vortices}
\label{sec:transfer-vortex}
\begin{figure}[tbp]
\centering
 \includegraphics[width=8cm]{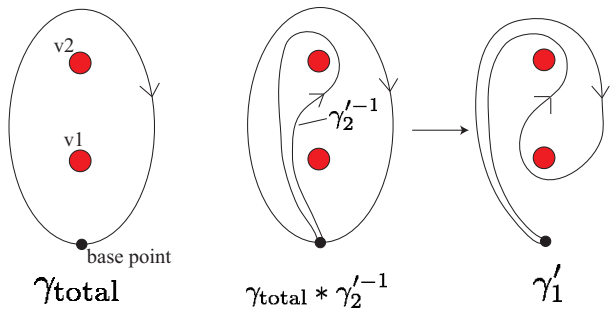}
 \caption{(color online) Schematic pictures of $\gamma_{\rm total}$ and $\gamma_1'= \gamma_{\rm total} \ast \gamma_2^{-1}$. In the rightmost figure, we show that $\gamma_1'$ is equivalent to a loop which encircles v1, but does not encircle v2.}\label{fig:conservation-vortex}
\end{figure}

Next, we relate the topological influence with the back-action via a charge transfer. Let us consider a situation in which two non-Abelian vortices with topological charges $\gamma_1$ and $\gamma_2 \in \pi_1 (G/H)$ coexist, as illustrated in the leftmost figure in Fig.~\ref{fig:conservation-vortex}. (See Fig.~\ref{fig:loop-deformation} for the definitions of $\gamma_1$ and $\gamma_2$.) Here we assume $\gamma_1$ does not commute with $\gamma_2$ and the total charge of the system is given by $\gamma_{\rm total} := \gamma_1 \ast \gamma_2$. As demonstrated in the previous section, the topological influence of $\gamma_1$ on $\gamma_2$ is given by
\begin{align}
 \gamma_2 \mapsto \gamma_2' : = \gamma_1^{-1} \ast \gamma_2 \ast \gamma_1, \label{eq:inf-NAvortex}
\end{align}
where $\gamma_2'$ is the topological charge after the topological influence. To make $\gamma_{\rm total}$ invariant under the topological influence, we require the topological back-action to satisfy
\begin{align}
 \gamma_1 \mapsto \gamma_1' := \gamma_{\rm total} \ast {\gamma_2'}^{-1}, \label{eq:back-NAvortex}
\end{align}
so that the topological charge conservation is met: $\gamma_1' \ast \gamma_2' = \gamma_1 \ast \gamma_2$. Here $\gamma_1'$ is depicted in the rightmost figure of Fig.~\ref{fig:conservation-vortex}.

As a concrete example, let us consider the biaxial nematic phase, where $\gamma_1$ and $\gamma_2$ are given by $i \sigma_x$ and $i \sigma_y$, respectively. From Eqs.~(\ref{eq:inf-NAvortex}) and (\ref{eq:back-NAvortex}), $\gamma_1'$ and $\gamma_2'$ are calculated to be $-i \sigma_x$ and $-i \sigma_y$. Thus, we can readily confirm the charge conservation.
In addition, Eqs.~(\ref{eq:inf-NAvortex}) and~(\ref{eq:back-NAvortex}) are rewritten as
\begin{align}
&\gamma_1' = \gamma_1 \ast [\gamma_1^{-1},\gamma_2]^{-1}, \\
&\gamma_2' = [\gamma_1^{-1} , \gamma_2] \ast \gamma_2,
\end{align}
where the commutator $[\gamma_1^{-1},\gamma_2]$ is defined by
\begin{align}
[\gamma_1^{-1} , \gamma_2] := \gamma_1^{-1} \ast \gamma_2 \ast  \gamma_1 \ast \gamma_2^{-1}.
\end{align}
Thus, the topological influence is related to the back-action via the commutator of the fundamental group $[\gamma_1^{-1},\gamma_2]$, which plays a central role in the charge transfer. 

\subsection{Charge transfer for topological excitations with higher dimensionality ($n \ge 2$)}
\label{sec:transfer-monopole}
\begin{figure}[tb]
\centering
 \includegraphics[width=8cm]{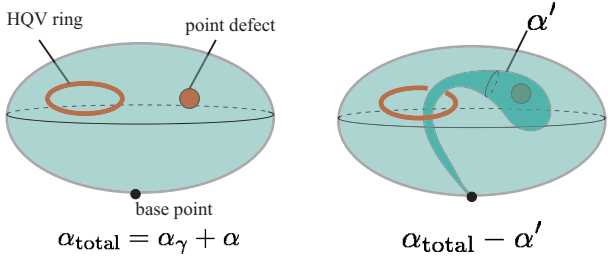}
 \caption{(color online) Schematic pictures of $\alpha_{\rm total}$ and $\alpha_{\rm total} - \alpha'$, where $\alpha_{\gamma}$ and $\alpha$ denote the topological charge of a half quantum vortex (HQV) and that of a point defect, respectively. In the right figure, we show a closed surface $\alpha_{\rm total} - \alpha'$ in which $\alpha'$ passes through a HQV and then encloses a point defect.}\label{fig:consevation-point}
\end{figure}

Similarly, we can define the charge transfer for the higher-dimensional case. We employ the method by Bucher {\it at el.}~\cite{Bucher:1992} for an Alice cosmic ring. We consider a system in which an HQV ring and a point defect coexist in 3D space as shown in the left figure of Fig.~\ref{fig:consevation-point}, where the topological charge of the HQV and that of the point defect are denoted by $\alpha_{\gamma}$ and $\alpha$ $\in \pi_2(G/H)$, respectively. In this situation, the total topological charge is given by $\alpha_{\rm total} = \alpha_{\gamma} + \alpha$, where ``$+$" means the additive operation since $\pi_n (G/H)$ ($n \ge 2$) is Abelian. According to Sec.~\ref{sec:influence}, the topological influence is given by    
\begin{align}
 \alpha \mapsto \alpha' = \gamma^{-1} \ast \alpha \ast \gamma. \label{eq:inf_point}
\end{align} 
In a manner similar to the case of non-Abelian vortices, we define the topological back-action as  
\begin{align}
 \alpha_{\gamma} \mapsto \alpha_{\gamma}' = \alpha_{\rm total} - \alpha'. \label{eq:back_point}
\end{align}
The image of $\alpha'_{\gamma}$ is illustrated in the right figure of Fig.~\ref{fig:consevation-point}.
Using a commutator of the homotopy groups, we can rewrite Eqs.~(\ref{eq:inf_point}) and~(\ref{eq:back_point}) as 
\begin{align}
&\alpha_{\gamma}' =  \alpha_{\gamma} - [\gamma^{-1},\alpha], \\
&\alpha' =  \alpha + [\gamma^{-1},\alpha],
\end{align}
where the commutator is defined by
\begin{align}
[\gamma^{-1}, \alpha] := \gamma^{-1} \ast \alpha \ast \gamma - \alpha.  
\end{align}
Thus, in the case of the point defect, the commutator $[\gamma^{-1},\alpha]$ describes the charge transfer due to the topological influence. For example, in the uniaxial nematic phase, the charge of a point defect changes from $\alpha$ to $-\alpha$ under the topological influence. Thus, the topological back-action is given by $\alpha_{\gamma}'= \alpha_{\gamma} + 2 \alpha$. Especially, for $\alpha_{\gamma} = 0$ and $\alpha = 1$, we have $\alpha_{\gamma}'= 2$, which corresponds to the twist of a vortex discussed in the previous subsection. Using the same approach, we can show that the charge transfer is generally classified by the commutator of homotopy groups $[\gamma^{-1},\beta]$, where $\gamma \in \pi_1 (G/H)$ and $\beta \in \pi_n (G/H)$. 

\subsection{Charge transfer in a multiple topological excitation system}
\label{sec:top_class}
 Finally, we make a general remark on the charge transfer for a multiple topological excitation system. As we discussed in previous subsections, the charge transfer indeed occurs between two topological excitations. In this subsection, consider a general situation in which there are $l$ topological excitations, and investigate the effect of the charge transfer between two of them on the remaining $l-2$ topological excitations. Let us define the total charge of $l$ non-Abelian vortices $\gamma_i$ ($l=1,2,\dots,l$) by
\begin{align}
\gamma_{\rm total} = \prod_{i=1}^{l} \; \gamma_{n_i},  \label{eq:tot_1dim}
\end{align}
where a set of labels $\{ n_i \}$ specifies the order of non-Abelian vortices which are placed from the left to the right with an order of increasing $n_i$. 
Also, the total charge of $l_1$ vortex rings $\alpha_{\gamma_{i}}$ ($i=1,\cdots l_1$) and $l_2$ point defects $\alpha_{j}$ ($j=1,\cdots ,l_2$) is defined by  
\begin{align}
\alpha_{\rm total} =  \left( \sum_{i=1}^{l_1}  \; \alpha_{\gamma_i} \right) + \left( \sum_{j=1}^{l_2} \; \alpha_{j} \right), \label{eq:tot_ndim}
\end{align}
where $l_1 + l_2 = l$. Under the topological charge conservation, we consider that a vortex with charge $\gamma_{n_j}$ rotates around a vortex with charge $\gamma_{n_i}$ (we assume $i < j$). By definition, the topological back-action and influence are given by 
\begin{subequations}
\begin{align} 
 &\gamma_{n_i} \mapsto \gamma_{n_i}' = \gamma_{n_i}\ast [\gamma_{n_i}^{-1}, \gamma_{n_j}]^{-1}, \label{eq:ex_back} \\
 &\gamma_{n_j} \mapsto \gamma_{n_j}' = [\gamma_{n_i}^{-1}, \gamma_{n_j}] \ast \gamma_{n_j}, \label{eq:ex_inf}
\end{align}
\end{subequations}
where the vortices labeled by $n_{i+1}, \cdots, n_{j-1}$ stay at their original positions. Although we do not perform any operations on the remaining vortices, the charge of vortices $\gamma_{n_{i+1}} \ast \cdots \ast\gamma_{n_{j-1}}$ changes so as to respect the conservation law:
\begin{align}
 &\gamma_{n_{i+1}} \ast \cdots \ast \gamma_{n_{j-1}} \notag \\
                &\mapsto [\gamma_{n_i}^{-1}, \gamma_{n_j}] \ast \gamma_{n_{i+1}} \ast \cdots \ast \gamma_{n_{j-1}} \ast [\gamma_{n_i}^{-1}, \gamma_{n_j}]^{-1}. \label{eq:ex_remain}
\end{align}
It follows from Eqs.~(\ref{eq:ex_back}), (\ref{eq:ex_inf}), and (\ref{eq:ex_remain}) that $\gamma_{\rm total}$ is invariant.  
Thus, the charge transfer from $\gamma_{n_i}$ to $\gamma_{n_j}$ gives an effect on the charge of vortices in the intermediate region. In general, the effect of a charge transfer depends on the initial condition of $\gamma_{\rm total}$. In other words, the topological charges are sensitive to a relative position of each vortex due to the noncommutativity of topological charges. In a similar fashion, when a point defect with charge $\alpha_j$ passes through a center of a vortex ring with charge $\alpha_{\gamma_i}$, the topological back-action and influence are given by 
\begin{subequations}
\begin{align}
 &\alpha_{\gamma_i} \mapsto \; \alpha_{\gamma_i}' = \alpha_{\gamma_i} - [\gamma_i^{-1}, \alpha_j], \label{eq:ex_back_ndim} \\
 &\alpha_j \mapsto \;  \alpha_j' =  \alpha_j +[\gamma_i^{-1}, \alpha_j]. \label{eq:ex_inf_ndim}
\end{align}
\end{subequations}
Since topological charges are Abelian, we can verify the charge conservation by substituting Eqs.~(\ref{eq:ex_back_ndim}) and (\ref{eq:ex_inf_ndim}) into Eq.~(\ref{eq:tot_ndim}). Thus, in the case of point defects, the charge transfer has no effect on other topological charges due to the commutativity. 

\section{Examples in Physical Systems}
\label{sec:examples}
In what follows, we show some concrete examples of charge transfers in liquid crystals, spinor BECs and superfluid $^3$He.

\subsection{Liquid crystal}
\label{sec:liquid_crystal}
As described in Sec.~\ref{sec:influence}, the topological influence on a non-Abelian vortex and a point defect exists for the uniaxial and biaxial nematic phase, respectively. Thus, these phases show a nontrivial charge transfer under the charge conservation. 

\subsubsection{Point defect in the uniaxial nematic phase}
Let us label a disclination and a point defect by $\gamma$ and $\alpha$, respectively. Then, the topological influence on the point defect is described by 
\begin{align}
 \gamma^{-1} \ast \alpha \ast \gamma = -\alpha.
\end{align}
Thus, by definition, the commutator is given by
\begin{align}
 [\gamma^{-1}, \alpha] = -2\alpha.
\end{align}
In general, a point defect with charge $n \in \mathbb{Z}$ is labeled by $n \alpha$, and thus $[\gamma^{-1}, n\alpha] = -2n\alpha$. Therefore, the charge transfer for the point defect is classified by 
\begin{align}
 [\pi_1, \pi_2] \cong 2 \mathbb{Z}. \label{eq:CT-UN}
\end{align}
where $``2\mathbb{Z}"$ implies even integers.

\subsubsection{Disclination in the biaxial nematic phase}
In the biaxial nematic phase, disclinations are classified by the eight-element quaternion group $Q_8$ in Eq.~(\ref{eq:vortex_BN}). The commutator of $i \sigma_x$ and $i \sigma_y$ gives
\begin{align}
 [i \sigma_x, i \sigma_y ] &= (i \sigma_x )(i \sigma_y) (i \sigma_x)^{-1} (i \sigma_y)^{-1} \notag \\
                           & =  (i \sigma_z )^2 = - \bm{1}_2.
\end{align}
 This result holds for all commutators $[i \sigma_{\mu}, i \sigma_{\nu}]$ with $\mu \neq \nu$. As $- \bm{1}_2 $ is its own inverse in the quaternion group, the action and back-action are identical. The charge transfer is classified by 
\begin{align}
 [Q_8,Q_8] \cong \mathbb{Z}_2.
\end{align}
Thus, there is only one nontrivial charge transfer.

\begin{table*}[tbp]
\centering
 \caption{Classification of charge transfers in condensed matter systems. The rightmost four columns show the classification of line defects ($\pi_1$), point defects ($\pi_2$), charge transfer between non-Abelian vortices ($[\pi_1,\pi_1]$), and charge transfer between a line defect and a point defect ($[\pi_1,\pi_2]$). For the sake of completeness, we also show the case with a trivial charge transfer; namely, the case in which  underlying homotopy groups are Abelian. \vspace*{2mm}} \label{tab:core}
 \begin{tabular}{ccccccc} \hline \hline
 System & Phase  & Order parameter manifold & $\pi_1$ &$\pi_2$ & $[\pi_1,\pi_1]$ & $[\pi_1,\pi_2]$ \\ \hline
 liquid crystal & UN & $\mathbb{R}$P$^2$\cite{Gennes:1993} & $\mathbb{Z}_2$ & $\mathbb{Z}$ & 0 & $2\mathbb{Z}$ \\
                & BN & $SO(3)/D_2$ \cite{Gennes:1993} & $Q_8$ & 0 & $\mathbb{Z}_2$ & 0 \\
  spin-1 BEC    & FM & $SO(3)_{\phi, \bm{f}}$\cite{Ho:1998} & $\mathbb{Z}_2$ &  0  & 0 & 0 \\
                &polar& $(U(1)_{\phi} \times S^2_{\bm{f}})/(\mathbb{Z}_2)_{\phi,{\bm{f}}}$\cite{Zhou:2001} & $\mathbb{Z}$ & $\mathbb{Z}$ & 0 & $2\mathbb{Z}$  \\ 
  spin-2 BEC    & F1 & $SO(3)_{2 \phi,\bm{f}}$\cite{Makela:2003,Makela:2006} & $\mathbb{Z}_4$ &  0  & 0 & 0 \\         
                & F2 & $SO(3)_{\phi,\bm{f}}$\cite{Makela:2003,Makela:2006} & $\mathbb{Z}_2$  &  0  & 0 & 0  \\
                & UN & $U(1)_{\phi} \times S^2_{\bm{f}}/(\mathbb{Z}_2)_{{\bm{f}}}$\cite{Song:2007,Uchino:2010} & $\mathbb{Z}$ & $\mathbb{Z}$ & 0 & $2\mathbb{Z}$  \\
                & BN & $(U(1)_{\phi} \times SO(3)_{\bm{f}})/(D_4)_{{\phi,\bm{f}}}$\cite{Song:2007,Uchino:2010} & $\mathbb{Z} \times_h (D_4^{\ast})_{{\phi,\bm{f}}}$ & 0 & $\mathbb{Z}_4$ & 0 \\ 
                & cyclic & $(U(1)_{\phi} \times SO(3)_{\bm{f}})/(T)_{{\phi,\bm{f}}}$ \cite{Makela:2003,Makela:2006,Semenoff:2007} & $\mathbb{Z} \times_h (T^{\ast})_{{\phi,\bm{f}}}$ & 0 & $Q_8$ & 0 \\
  $^3$He-A      & dipole free  & $(SO(3)_{\bm{L}} \times S^2_{\bm{S}})/(\mathbb{Z}_2)_{\bm{S},\bm{L}}$\cite{Volovik:2003} & $\mathbb{Z}_4$ & $\mathbb{Z}$ & 0 & $2 \mathbb{Z}$ \\ 
                & dipole locked & $SO(3)_{\bm{S},\bm{L}}$\cite{Volovik:2003} & $\mathbb{Z}_2$ & 0 & 0 & 0 \\
                
  $^3$He-B      & dipole free & $U(1)_{\phi} \times SO(3)_{\bm{S}+\bm{L}}$\cite{Volovik:2003} & $\mathbb{Z} \times \mathbb{Z}_2$ & 0 & 0 & 0 \\  
                & dipole locked & $U(1)_{\phi} \times S^2_{\bm{S}+\bm{L}}$\cite{Volovik:2003} & $\mathbb{Z}$ & $\mathbb{Z}$ & 0 & 0 \\   
               
 \hline \hline
 \end{tabular}
\end{table*}

\subsection{Spinor Bose-Einstein Condensates}
Next, we see the charge transfer for spin-1 and spin-2 BECs. The spin-1 BECs have been realized for $^{23}$Na and $^{87}$Rb and the spin-2 BEC for $^{87}$Rb. According to the homotopy classification in Refs.~\cite{Kobayashi:2012,Kawaguchi:2012,Makela:2003,Makela:2006,Semenoff:2007}, non-Abelian vortices exist for spin-2 biaxial nematic and spin-2 cyclic phases. Also, the topological influence on a point defect arises for the spin-1 polar phase and the spin-2 uniaxial nematic phase. Since the charge transfer of point defects is calculated in a manner similar to the case of the uniaxial nematic phase, we consider here only the non-Abelian vortices.   

\subsubsection{Vortex in the spin-2 biaxial nematic phase}
The OPM of the biaxial nematic phase of a spin-2 BEC is given by \cite{Song:2007,Uchino:2010}
\begin{align}
 G/H_{\rm Spin2 \; BN} &= (U(1)_{\phi} \times SO(3)_{\bm{f}})/(D_4)_{\phi,\bm{f}} \notag \\
                                &= (U(1)_{\phi} \times SU(2)_{\bm{f}})/(D^{\ast}_4)_{\phi,\bm{f}}, 
\end{align}
where the subscripts $\phi$ and $\bm{f}$ indicate the global gauge symmetry and the spin rotation symmetry, respectively, $(D_4)_{\phi,\bm{f}}$ is a spin-gauge-coupled fourth dihedral group, and $(D^{\ast}_4)_{\phi,\bm{f}}$ is its binary group due to the lift of $SO(3)$. The fundamental group is given by \cite{Kobayashi:2012}
\begin{align}
 \pi_1 ( G/H_{\rm Spin2 \; BN}) \cong \mathbb{Z} \times_h (D_4^{\ast})_{\phi,\bm{f}}, \label{eq:GH_SBN}
\end{align}
where we introduce an $h$-product $``\times_h"$ since $(D_4^{\ast})_{\phi,\bm{f}}$ involves the global gauge degrees of freedom. The definition of the $h$-product is given in Appendix~\ref{app:A}.

Since $\mathbb{Z}$ is Abelian, we can ignore it and we discuss only the binary fourth dihedral group $(D_4^{\ast})_{\phi,\bm{f}}$. The fourth dihedral group consists of the identity element, three four-fold rotations, and a two-fold rotation about an axis perpendicular to the rotation axis. Since the binary fourth dihedral group is the double representation of $(D_4)_{\phi,\bm{f}}$, it can simply be described by
\begin{align}
(D_4^{\ast})_{\phi,\bm{f}} \cong \langle r,s | r^8 = 1, s^2 = r^4, s^{-1} r s = r^{-1} \rangle, \label{eq:equiv_form_D4}
\end{align} 
where $r$ and $s$ are the generators of the group satisfying the three relations on their right. Then, an element of $(D_4^{\ast})_{\phi,\bm{f}}$ is described by $r^n$ and $r^n s $ ($n=1,2,\cdots 8$). The nontrivial commutators arise from $[r^n, r^m s]$ and $[r^n s , r^m s]$ ($n \neq m$), which are calculated to be
\begin{subequations}
\begin{align}
 [r^n, r^m s] &=  r^{2n}, \label{eq:calc_rrs}\\ 
 [r^n s, r^m s] &=  r^{2(n-m)}. \label{eq:calc_rsrs}
\end{align}
\end{subequations}
In Eq.~(\ref{eq:calc_rrs}), $m$ has no effect on the commutator because we have $[r^n, r^m s] = r^{n+m}s r^{-n}s^{-1} r^{-m}$ $=r^{n+m}r^{n-m}$. Therefore, the commutator subgroup is given by
\begin{align}
 [(D_4^{\ast})_{\phi,\bm{f}},(D_4^{\ast})_{\phi,\bm{f}}] \cong \mathbb{Z}_4.
\end{align}
This result implies that we have only three nontrivial charge transfers.
Furthermore, in this case, we can immediately see that some pairs of vortices behave as $\mathbb{Z}_2$. In fact, considering winding an $r^m s$ vortex around an $r^n$ vortex, we find that the commutator in this case is given in Eq.(\ref{eq:calc_rrs}) as $r^{2n}$. Thus, the topological back-action becomes 
\begin{align}
 r^n \mapsto r^n [r^n,r^m s]^{-1} = r^{-n},
\end{align}
which implies that a second winding will reverse this action. We therefore have a $\mathbb{Z}_2$-action. However, if we consider winding an $r^m s$ vortex around an $r^n s$ vortex, the commutator is $r^{2(n-m)}$, and therefore
\begin{subequations}
\begin{align}
 r^m s \mapsto [r^ms,r^ns]r^m s &= r^{m+2(n-m)}s \label{eq:calc_rm-rn_a},\\
 r^n s \mapsto r^n s [r^ms,r^ns]^{-1}&= r^n s r^{-2(n-m)} \notag \\
                                &= r^n s r^{-2(n-m)}s^{-1} s \notag \\
                                &= r^{n+2(n-m)}s. \label{eq:calc_rm-rn_b}
\end{align} 
\end{subequations}
Equations~(\ref{eq:calc_rm-rn_a}) and (\ref{eq:calc_rm-rn_b}) are simply expressed by
\begin{subequations}
\begin{align}
& m \mapsto m' = m+2(n-m) \\
& n \mapsto n' = n+2(n-m) 
\end{align}
\end{subequations}
Thus, we find $m'-n' = m-n$. Since $m,n \in \mathbb{Z}_8$, we see immediately that this will be a $\mathbb{Z}_4$-action if $n-m$ is odd, and a $\mathbb{Z}_2$-action if $n-m$ is even. 

\subsubsection{Vortex in the spin-2 cyclic phase}
Next, we consider the cyclic phase of a spin-2 BEC, whose OPM is given by~\cite{Makela:2003,Makela:2006,Semenoff:2007}
\begin{align}
 G/H_{\rm C} &= (U(1)_{\phi} \times SO(3)_{\bm{f}})/T_{\phi,\bm{f}} \notag \\
                                &= (U(1)_{\phi} \times SU(2)_{\bm{f}})/T^{\ast}_{\phi,\bm{f}},
\end{align}
where $T_{\phi,\bm{f}}$ is a spin-gauge-coupled tetrahedral group and $T^{\ast}_{\phi,\bm{f}}$ is its binary form. The fundamental group becomes \cite{Kobayashi:2012}
\begin{align}
 \pi_1 ( G/H_{\rm C}) \cong \mathbb{Z} \times_h T^{\ast}_{\phi,\bm{f}}, \label{eq:GH_C}
\end{align}
where $\mathbb{Z}$ results from the $U(1)$ degrees of freedom. Since $T^{\ast}_{\phi,\bm{f}}$ is the spin-gauge coupled symmetry, $\mathbb{Z}$ and $T^{\ast}_{\phi,\bm{f}}$ are described by the $h$-product defined in Appendix~\ref{app:A}. Similar to the biaxial nematic phase, we can ignore $\mathbb{Z}$, since $\mathbb{Z}$ is Abelian. We consider the commutator of $T^{\ast}_{\phi,\bm{f}}$. We can show the equivalent form of $T^{\ast}_{\phi,\bm{f}}$ as
\begin{align}
 T^{\ast}_{\phi,\bm{f}} \cong \mathbb{Z}_3 \ltimes_c Q_8, \label{eq:T_equivform}
\end{align} 
where the semidirect product ``$\ltimes_c$'' is defined as follows: element $\bm{0}$ in $\mathbb{Z}_3=\{\bm{0},\bm{1},\bm{2}\}$ does nothing; element $\bm{1}$ performs a permutation, i.e., $i \sigma_x \to i \sigma_y$, $i \sigma_y \to i \sigma_z$, and $i \sigma_z  \to i \sigma_x$; and element $\bm{2}$ performs the cyclic permutation, i.e., $i \sigma_x \to i \sigma_z$, $i \sigma_y \to i \sigma_x$, and $i \sigma_z \to i \sigma_y$. For example, we consider the product of $(\bm{1}, i \sigma_x)$ and $(\bm{1}, i \sigma_y) \in \mathbb{Z} \ltimes_c Q_8$, which is given by
\begin{align}
 (\bm{1}, i \sigma_x)(\bm{1}, i \sigma_y) &= (\bm{2}, (i\sigma_y) (i \sigma_y)  ) \notag \\
             & = (\bm{2}, - \bm{1}_2).
\end{align} 
To derive the commutator subgroup, we consider the following two commutators:
\begin{subequations}
\begin{align}
 [(\bm{0},i \sigma_x), (\bm{0}, i \sigma_y)] &= (\bm{0},i \sigma_x) (\bm{0}, i \sigma_y)(\bm{0},-i \sigma_x) (\bm{0}, -i \sigma_y ) \notag \\
                                       &= (\bm{0}, -\bm{1}_2), \\
  [(\bm{1},i \sigma_x), (\bm{1}, i \sigma_y)] &= (\bm{1},i \sigma_x) (\bm{1}, i \sigma_y)(\bm{2},-i \sigma_z) (\bm{2}, -i \sigma_x ) \notag \\
                                       &= (\bm{0}, i \sigma_z),                                       
\end{align}
\end{subequations}
where the first expression indicates that we have the $-\bm{1}_2$ element of $Q_8$, and the second one shows that $i \sigma_x$, $i\sigma_y$, and $i\sigma_z$ belong to the commutator subgroup. The closure of the commutator subgroup implies that it is at least the quaternions. However, as the $\mathbb{Z}_3$ is Abelian, the commutator subgroup is isomorphic to the quaternion; namely, 
\begin{align}
 [T^{\ast}_{\phi,\bm{f}},T^{\ast}_{\phi,\bm{f}}] \cong Q_8.
\end{align}
Thus, the charge transfer in the cyclic phase is described by the eight-element quaternion group $Q_8$, and furthermore since $Q_8$ is non-Abelian, the cyclic phase accommodates the non-Abelian charge transfer.

\subsection{Superfluid $^3$He}
Finally, we show the existence of the topological back-action in the superfluid $^3$He, in which the order parameter possesses the orbital and spin degrees of freedom, and each internal degree of freedom behaves like a vector in the orbital and spin space due to the spin triplet $p$-wave state. In general, the order parameter is described by 
\begin{align}
\Delta_{\mu} = \sum_j d_{\mu j} \hat{k}_j,
\end{align}
where $\mu$ refers to $\mu= x,y,z$ in the spin space and $j$ refers to $j=x,y,z$ in the orbital space, and $d_{\mu j}$ is a complex 3-by-3 matrix. The possible phases and topological excitations are discussed in Refs.~\cite{Vollhardt:1990,Volovik:2003}. According to them, the topological influence between the vortex and the point defect exists for the dipole-free state in the superfluid $^3$He-A phase. Thus, we expect the topological back-action through the topological charge conservation. The order parameter is given by
\begin{align}
d_{\mu j} = \Delta_{\rm A} \hat{d}_{\mu} (\hat{m}_{j} + i \hat{n}_j), 
\end{align}  
where $\Delta_{\rm A}$ is an amplitude of the order parameter, and $\hat{\bm{d}}$, $\hat{\bm{m}}$, and $\hat{\bm{n}}$ are unit vectors which satisfy $\hat{\bm{m}} \perp \hat{\bm{n}}$. Also, the OPM is given by~\cite{Volovik:2003}
\begin{align}
G/H_{\rm A} = (SO(3)_{\bm{L}} \times S^2_{\bm{S}})/(\mathbb{Z}_2)_{\bm{S},\bm{L}},
\end{align}
where the subscripts $\bm{L}$ and $\bm{S}$ represent the orbital and spin symmetries. The corresponding first and second homotopy groups are given by
\begin{subequations}
\begin{align}
 &\pi_1(G/H_{\rm A}) \cong \mathbb{Z}_4, \\
 &\pi_2(G/H_{\rm A}) \cong \mathbb{Z}.
\end{align}
\end{subequations}
The topological influence occurs for $a$ and $a^3 \in \mathbb{Z}_4$, which give the $\mathbb{Z}_2$-action on $\pi_2$; i.e. $\alpha \in \pi_2$ changes into its inverse $- \alpha$. Therefore, the calculation of the charge transfer is the same as in the case of the uniaxial nematic phase. We summarize the classification of charge transfers in Table.~\ref{tab:core}.

\section{Testing of topological charges under the charge conservation}
\label{sec:testing}
In this section, we discuss the issues concerning the definition of the total charge and the measurement of topological charges. If the homotopy group is Abelian, we can observe topological excitations individually. In this case, the total charge is given by the sum of individual topological charges. The representative examples are superfluid $^4$He and scalar BECs (fully polarized spinor BECs), in which the order parameter manifold is given by $U(1)$. Since $\pi_1 (U(1)) \cong \mathbb{Z}$, it is possible to observe the winding number of the vortices individually. However, if the homotopy group is non-Abelian, the counting of individual defects breaks down due to the influence of the coexisting defects. In this case, the topological charge is determined up to the conjugacy class of the homotopy group. Thus, the sum of each individual topological charge is insufficient to define the total charge. Nevertheless, in the closed system, the total charge would be defined by encoding the values of the order parameter on the boundary of the system, which remains unchanged under the adiabatic manipulation. Thus, the total charge is measurable by probing the system boundary without regard to the topological influence. 

Next, we state the detection of each individual topological charge in the topological charge conserving framework. We suggest that we could determine individual topological charges through the testing of pair annihilation. For example, let us consider the system with a vortex v$_1$. Assuming that we can create a vortex-antivortex pair and manipulate vortices adiabatically, we create the vortex-antivortex pair: v$_2$ and v$_2$' near the vortex v$_1$, and then we rotate the vortex v$_2$ about the vortex v$_1$ adiabatically in the clockwise direction. Finally, we fuse the vortex v$_2$ into the remaining antivortex v$_2$'. The combined charge of v$_2$ and v$_2$' turns out to be given by a commutator. As a result, this test is categorized into three cases: (1) The vortex pair annihilates each other. Hence, each topological charge is Abelian and we can observe the topological charge individually. (2) The vortex pair survives and its charge belongs to the Abelian group. Then, the vortices v$_1$ and v$_2$ are not observable individually, but each individual commutator is observable. Thus, a relative charge of v$_1$ and v$_2$ is possible to detect through the commutator. (3) The vortex pair survives and its charge belongs to the non-Abelian group. In this case, it is also possible to determine the topological charge uniquely if we have only two types of the vortex. However, if there are more than two, we cannot determine the topological charge uniquely due to the influence between commutators. Namely, we can detect the charge of vortex-antivortex pair up to the conjugacy class.
Fortunately, for the real system tabulated at Table~\ref{tab:core}, the commutator subgroup is Abelian except for the cyclic phase in the spin-2 BECs. Thus, almost all ordered phases are categorized into the case (1) or (2). In the cyclic phase, the commutator subgroup is isomorphic to the eight-element quaternion group. Namely, this phase belongs to the case (3). 

As a concrete example, if we consider non-Abelian vortices in the spinor BECs, the total charge would be estimated by analyzing the spatial variations of the order parameter at the boundary through a Stern-Gerlach separation, in which the populations in each magnetic sublevel can be measured~\cite{Choi}.  On the other hand, each individual topological defect is determined up to the conjugacy class due to the topological influence. To observe the topological charge in the charge conserving scheme, it is necessary to apply the pair annihilation testing to individual topological defects. However, such manipulation of the topological defects has yet to be achieved in the experiment of spinor BECs.

\section{Summary and Discussion}
\label{sec:summary}
In the present paper, we have discussed the relation between the topological charge conservation and the topological influence. To ensure the consistency with the charge conservation, we have introduced the topological back-action, and shown that the topological back-action causes twisting of a vortex line for the point-defect case. The twist of a vortex corresponds to two HQV rings (Alice rings) attached on two edges of a vortex line, whose topological charge is characterized by $\pi_2 (G/H)$. This idea might be generalized to higher-dimensional cases ($n \ge 3$), in which a dimension of a vortex is $\nu = d-2$, where $d$ is the space dimension. Thus, a surface enclosing a vortex is given by $L (s_1, \cdots, s_{d-1}) \in \Omega^{d-1} (G/H)$, where $\Omega^n (G/H) = \Omega (\Omega^{n-1}(G/H)) $. Here, $L(s_1, \cdots, s_{d-1})$ includes the degrees of freedom to encircle the vortex core and we assume that each edge has the same configuration. The topological back-action from a point defect in a $d$-dimensional space is given by the homotopy equivalence classes of $L(s_1, \cdots, s_{d-1})$; i.e. 
\begin{align}
 \pi_0 (\Omega^{d-1} (G/H)) \cong \pi_{d-1} (G/H),
\end{align}
where $\pi_{d-1} (G/H)$ describes a topological charge of a point defect in the $d$-dimensional space. 

Using the method by Bucher {\it et al.}~\cite{Bucher:1992}, we have related the topological influence with its back-action via the commutators of the homotopy groups $[\pi_1,\pi_n]$ ($n \ge 1$), which plays a key role in the charge transfer in a  multiple topological excitation system. For liquid crystals, spinor BECs, and superfluid $^3$He, we have calculated commutators and determined the classifications of the charge transfer for the non-Abelian vortices: one type for the BN phase, three types for the spin-2 biaxial nematic phase, and seven types for the cyclic phase. In particular, we have found non-Abelian charge transfer in the cyclic phase, which may lead to new topological phenomena. It would be worthwhile to mention that the commutator between $\pi_1$ and $\pi_n$ represents the Whitehead bracket \cite{Whitehead}, which suggests a new direction of this study. 

In higher-dimensional cases, the UN phase, the spin-1 polar phase, the spin-2 uniaxial phase, and the superfluid $^3$He-A phase show the charge transfer classified by an even integer due to the topological influence of the HQV (disclination). This charge transfer depends mainly on a set of the automorphism maps $\aut(\pi_n)$. Also, if we consider the action of $\pi_1$ on $\pi_n$ to be a homomorphism from $\pi_1$ to $\aut( \pi_n )$, then we can rule out topological influence in a number of different cases on group-theoretic grounds. In particular, when $\pi_2 \cong \mathbb{Z}$, $\aut(\mathbb{Z})$ is always $\mathbb{Z}_2$. Thus, we can have only the $\mathbb{Z}_2$-action on the point defect. For example, in the case of the dipole-free $^3$He-A phase, $\pi_1(G/H) = \mathbb{Z}_4$ and $\pi_2(G/H) = \mathbb{Z}$, and the $\pi_1$ action on $\pi_2$ is nontrivial~\cite{Volovik:1977inf}. For $\mathbb{Z}_4 \cong \{ 1,a,a^2,a^3 \}$, if we consider the homomorphism map from $\mathbb{Z}_4$ into $\aut( \mathbb{Z} ) = \mathbb{Z}_2$, the only such homomorphism takes $a^2$ to an identity map. Thus, $1$ and $a^2$ will not cause any effect, while $a^1$ and $a^3$ will.

Also, we have suggested that the pair annihilation testing enables us to detect the topological charge in the charge conservation framework. This testing is applicable to the higher dimensional cases and is useful for detecting evidence of the nature of non-Abelian charge.   

In this paper, we have focused on the role of the charge conservation in a multiple topological excitation system. However, our calculation is also applicable to statistics of topological excitations for situations in which two topological excitations exchange their charges. It has been shown in Refs.~\cite{Lo:1993,Brekke:1991,Brekke:1993} that non-Abelian vortices obey the ``partially colored braid statistics" because a topological charge of a vortex is indistinguishable within the conjugacy class. The biaxial nematic phase in the liquid crystal exhibits the non-Abelian vortex (line defect) classified by $Q_8$, but its commutator is $\{\pm \bm{1}_2\}$. Thus, the exchange of vortices has a simple algebraic structure. On the other hand, the spin-2 biaxial nematic and spin-2 cyclic phases exhibit several different types of the charge transfer, offering a piece of evidence on the richness of statistics.    

\section*{ACKNOWLEDGMENTS}
This work was supported by KAKENHI 22340114, a Grant-in-Aid for Scientific Research on Innovation Areas ``Topological Quantum Phenomena" (KAKENHI 22103005) and the Photon Frontier Network Program, from MEXT of Japan. S.K. acknowledges support from JSPS (Grant No. 228338).  

\appendix

\section{The definitions of $\rtimes$ and $\times_h$} 
\label{app:A}
In this appendix, we give the definitions of $\rtimes$ and $\times_h$, which are used in Eqs. (\ref{eq:GH_UN}), (\ref{eq:GH_SBN}), and (\ref{eq:GH_C}) to make this paper self-contained.    
\subsection{The semidirect product}
The semidirect product of two groups $G_1$ and $G_2$, which is denoted as $G_1 \rtimes G_2$, is defined as follows; for $g_1,g_1' \in G_1$ and $g_2,g_2' \in G_2$, the semidirect product between $(g_1,g_2)$ and $(g_1',g_2')$ is given by
\begin{align}
(g_1,g_2) (g_1',g_2') = (g_1 f_{g_2} (g_1')  , g_2 g_2'),     
\end{align}
where $f_{g_2} \in \aut (G_1)$. For instance, in Eq.~(\ref{eq:GH_UN}), $G_1$ and $G_2$ are given by $ SO(2) \cong U(1)$ and $\mathbb{Z}_2 = \{\bm{0}, \bm{1}\}$, respectively. For $e^{i \theta}, e^{i \theta'} \in U(1)$ and $g,g' \in \mathbb{Z}_2$, the semidirect product is given by
\begin{align}
 (e^{i \theta},g) (e^{i \theta'},g' ) = (e^{i \theta} f_{g} (e^{i \theta'}) ,g+g'),  
\end{align}
where $f_{g}$ is given by
\begin{align}
 f_{g} (e^{i \theta}) = \begin{cases} e^{-i\theta} \; &\text{if } g = \bm{1}; \\ e^{i\theta} \; &\text{if } g = \bm{0}. \end{cases}
\end{align}

\subsection{The $h$-product}
To satisfy the group structure between the spin-gauge-coupled discrete group $K_{\phi, \bm{f}}$ and $\mathbb{Z}$, we define an $h$-product $\times_h$ as follows: given a discrete group $K_{\phi,\bm{f}}$ involving both the global gauge and the spin rotation, the group structure of $\mathbb{Z} \times_h K_{\phi,\bm{f}}$ is given by 
\begin{align}
(k, g)(k' ,g') = (k+k' + h(g,g'), e^{- i2 \pi h(g,g')}gg')
\end{align}
where $k,k' \in \mathbb{Z}$ and $g= (e^{i \theta},g_n)$, $g' = (e^{i \theta'},g_n')$, and $gg' = (e^{i (\theta+\theta')},g_ng_n') \in K_{\phi,\bm{f}}$. Here $h$ is a map from $K_{\phi,\bm{f}} \times K_{\phi,\bm{f}}$ to $\mathbb{Z}$, which is given by
\begin{align}
h(g,g') = \begin{cases} 0 \; &\text{if }\theta + \theta' < 2 \pi; \\ 1 \; &\text{if }\theta + \theta'  \ge 2 \pi. \end{cases}
\end{align}
Examples for the spin-2 biaxial nematic and spin-2 cyclic phases are described in Ref.~\cite{Kobayashi:2012}.

\end{document}